\documentclass[twocolumn]{aastex63}

\newcommand{\OmK}{\Omega_\mathrm{K}}
\newcommand{\st}{S_\mathrm{t}}
\newcommand{\wdust}{w_\mathrm{d}}
\newcommand{\wdustone}{w_\mathrm{d1}}
\newcommand{\wdusttwo}{w_\mathrm{d2}}
\newcommand{\wgas}{w_\mathrm{g}}

\newcommand{\sigg}{\Sigma_\mathrm{g}}

\date{Edited \today}
\accepted{July 24, 2020}
\submitjournal{ApJ}

\shorttitle{Disk-Planet Neural Network}
\shortauthors{Auddy \& Lin}
\graphicspath{{./}{figures/}}

\begin{document}

\title{A Machine Learning model to infer planet masses from gaps observed in protoplanetary disks}

\correspondingauthor{Sayantan Auddy}
\email{sauddy@asiaa.sinica.edu.tw}

\author[0000-0003-3784-8913]{Sayantan Auddy}
\affil{Institute of Astronomy and Astrophysics, Academia Sinica, Taipei 10617, Taiwan}

\author[0000-0002-8597-4386]{Min-Kai Lin}
\affiliation{Institute of Astronomy and Astrophysics, Academia Sinica, Taipei 10617, Taiwan}










\begin{abstract}
Observations of bright protoplanetary disks often show annular gaps in their dust emission. One interpretation of these gaps is disk-planet interaction. If so, 
fitting models of planetary gaps to observed protoplanetary disk gaps can reveal the presence of hidden planets. However, future surveys are expected to produce an ever-increasing number of protoplanetary disks with gaps. In this case, performing a customized fitting for each target becomes impractical owing to the complexity of disk-planet interaction. To this end,{ we introduce DPNNet (Disk Planet Neural Network)}, an efficient model of planetary gaps by exploiting the power of machine learning. {We train a deep neural network} with a large number of dusty disk-planet hydrodynamic simulations across a range of planet masses, disk temperatures, disk viscosities, disk surface density profiles, particle Stokes numbers, and dust abundances. The network can then be deployed to extract the planet mass for a given gap morphology. In this work, first in a series, we focus on the basic concepts of our machine learning framework. We demonstrate its utility by applying it to the dust gaps observed in the protoplanetary disk around HL Tau at $10$ au, $30$ au, and $80$ au. Our network predict planet masses of $80 \, M_\Earth$,  $63 \, M_\Earth$, and $70 \, M_\Earth$, respectively, which are comparable to other studies based on specialized simulations. We discuss the key advantages of our {DPNNet} in its flexibility to incorporate new physics, any number of parameters and predictions, and its potential to ultimately replace hydrodynamical simulations for disk observers and modelers. 

\end{abstract}


\section{Introduction} \label{sec:intro}
In the past few decades, exoplanet surveys using multiple techniques have revealed that planets are essentially ubiquitous throughout the galaxies \citep{cas12,bat13}. Their distribution in sizes and diversity in the masses, radii, and composition \citep[e.g.,][]{win15,ful17} have put constraints over planet formation theories \citep[see review by][]{joh14,ray20}. However, with the current planet search methods \citep{fis14}, it is often difficult to detect planets around young stars. Planet signatures like spectra and/or light curve from such systems are faint due to enhanced stellar activity in young stars and the presence of protoplanetary disk (hereafter PPDs). Thus, most discovered exoplanets are old ($\sim 10^3 \rm {Myr}$), with few detection of young planet candidates in systems $<10 \rm {Myr}$ old \citep[e.g.,][]{yu17,kep18}.  This limits our ability to constrain the demographics of young planets, particularly during their formation epoch in PPDs. An alternative and/or indirect method to probe the unseen younger population of exoplanets is needed as that is key to testing planet formation theories.

\begin{figure}[ht]
\centering
\includegraphics[width=3.6in,trim=50mm 40mm 50mm 35mm, clip=True]{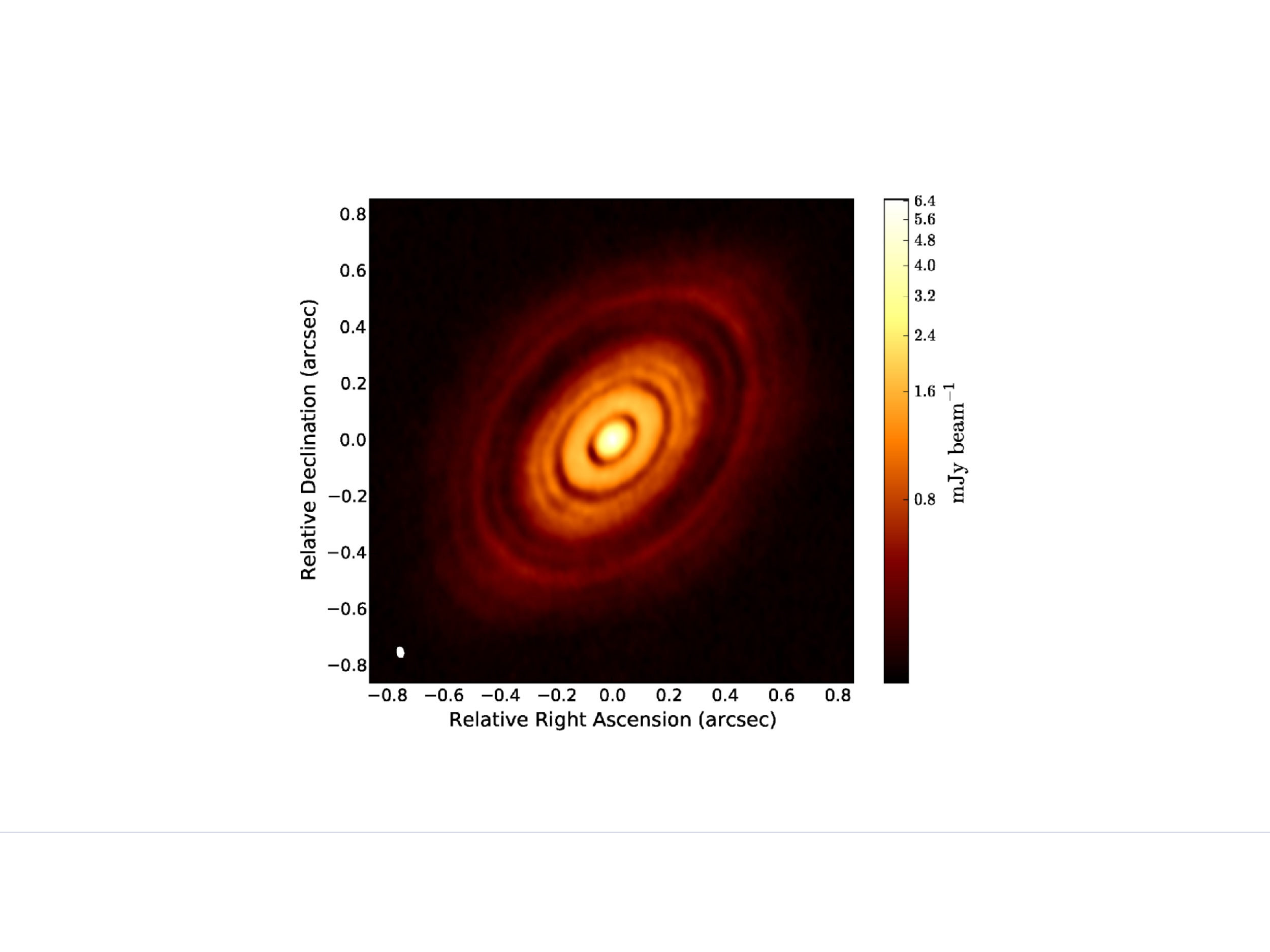}
\caption{{ ALMA continuum image of HL Tau \citep{ALMA15}. The synthesized beam is shown in the lower left. Adapted from \cite{Dipierro2015OnTau}.}}
\label{fig:HL-Tau_ALMA}
\end{figure}

In this work, we introduce a novel state-of-the-art scheme that uses artificial intelligence (machine learning techniques) to infer properties of young, undetected planets from observed features in PPDs. Our idea is motivated by recent high resolution observations  \citep[e.g.,][]{Clarke2018High-resolutionAu,liu19,per18,Ands18,Hua18b,Huac,Long2018GapsRegion} revealing complex -- possibly planet-induced -- features in PPDs with unprecedented precision. Near-infrared imaging with Atacama Large Millimeter Array (ALMA) has published a plethora of resolved images of PPDs  showing detailed substructures like rings and distinct gaps in systems like HL Tau,{ see Figure} (\ref{fig:HL-Tau_ALMA}), \citep{ALMA15,yen16}, HD 169141  \citep{mom15}, HD 97048 \citep{van17}, and TW Hya \citep{and16,Hua18a}.

The origin of such rings and gaps, although debated extensively, are broadly categorized into those caused due to disk physics and chemistry and those due to disk-planet interaction. Disk-specific mechanism include zonal flows  associated with magnetorotational instability (MRI) turbulence  \citep[e.g.][]{joh09,sim14}, secular gravitational instability \citep{you11,tak14}, condensation of molecular species leading to growth and fragmentation of dust around various snowlines  \citep[e.g.,][]{Zhang2015EVIDENCEDISK, Pinilla2017DustDisks}, self-induced dust pile-ups \citep{gon15}, gap-opening by 
large-scale vortices \citep{bar17}, etc.  

The alternative explanation, and the one that is of interest to us, is that these gaps are induced by embedded young planets due to its dynamical interaction with the disk \citep[][]{Rosotti2016TheObservationsb,Dipierro2016TwoDiscs,Dong2017WhatPlanet}. Although disk-planet interaction has long been predicted, \citep{goldreich80,lin93}, direct observational evidence has only come recently \citep{pinte19,pinte20}, which relies on careful kinematic measurements. On the other hand, annular dust gaps are readily observed global features and thus offer a promising indirect method to detect planets. To this end, a model constraining planet properties from gap features is key to the exploration of the unseen young exoplanet population.

It should be noted that gap and ring formation due to planet-disk interaction and that due to disk processes are not mutually exclusive. For example, snow lines can favour planetesimal formation \citep{drazkowska17} by acting as dust traps. In this case, ring formation at the snow line \citep{morbi20} results in planet formation, which can then induce planet gaps and rings. It is also possible that planet and non-planet gap/ring formation processes occur in same disk at different radii. However, including additional gap/ring formation processes is beyond the scope of this work. Our model assumes that  planet-disk interaction dominates over other mechanisms.

\subsection{Identifying young planets from gap profiles}

In this paper, we focus on predicting planet masses from observed, axisymmetric disk gaps in dust emission. 
For each gap, there are at least two measurable parameters: its depth and  width. The depth is the contrast between the undepleted background and the minimum intensity inside the gap, and the width is defined as the radial extent between the inner and outer edge of the gap \citep[for example see][]{Dong2017WhatPlanet,Clarke2018High-resolutionAu,Long2018GapsRegion}

Both gap depths and widths have been modeled  extensively using analytical and numerically approaches  \citep[e.g.][]{crida06, paardekooper09, Duffell2013GAPDISK, Fung2014HOWPLANETS,Duffell2015ADISKS,Kanagawa2015MASSSTRUCTURES,Kanagawa2016MassWidth}. However, instrumental limitation (i.e., lack of sensitivity resulting in poor signal-to-noise ratio) often makes it hard to constrain gap depths observationally \citep{Zhang2018TheInterpretation}. Thus, gap widths are often considered a better alternative to constrain planet mass \citep{Duffell2013GAPDISK}. 

Recently, \citet{Lodato2019TheDiscs} used a simple empirical relation (hereafter the Lodato model) to infer planet masses from observed gap widths. They assumed the dust gap width $\wdust = k R_{\rm H} $,{ where $k$ is the proportionality constant} and $R_\mathrm{H} = (M_{\rm  P}/3M_{*})^{1/3}R_0$ is the planet's Hill radius, $R_0$ is its orbital radius, and $M_{\rm P,*}$ are the planet and stellar masses, respectively. This gives

\begin{equation}
    M_\mathrm{P} = \left(\frac{\wdust}{k R_0}\right)^3 3 M_*.\label{lodato_model}
\end{equation}
{\cite{Lodato2019TheDiscs} assumed a value of $k=5.5$ by averaging results from hydrodynamical simulations of CI Tau and MWC 480.}
While the simplicity of this model is appealing, it does not account for the fact that gap profiles also depend on disk properties \citep{crida06,paardekooper09}.

A more sophisticated approach was taken by \cite{Kanagawa2016MassWidth}. They ran multiple two-dimensional (2D) hydrodynamic simulations to model gap opening in gaseous disks for relatively massive planets $(0.1 M_{\rm J}- 2M_{\rm J})$ and derived an empirical formula (hereafter the Kanagawa model)  
\begin{equation}
    M_{\rm P}=  0.0021 \times \left(\frac{\wgas}{R_{\rm 0}}\right)^2 \left(\frac{h_{\rm 0}}{0.05}\right)^{\frac{3}{2}} \left(\frac{\alpha}{10^{-3}}\right)^{\frac{1}{2}}M_* 
\label{eq:kanagawa}
\end{equation}

to relate planet masses and {\it gas} gap widths, $\wgas$. This relation also depends on the disk's local aspect-ratio $h_0$ and dimensionless viscosity parameter $\alpha$.

A limitation of Kanagawa model is that it requires gap widths in gas emission,  which do not coincide with observed dust gaps unless particles are strongly coupled to the gas \citep{Dipierro2016TwoDiscs}. Moreover, low mass planets can open up dust gaps only, with negligible effect on the gas profile  \citep{Dipierro2016TwoDiscs,Rosotti2016TheObservationsb}. As will be shown later, using dust gaps instead of gas gaps in Eq. \ref{eq:kanagawa} leads to an overestimation of planet masses when applying the Kanagawa model. 
 
In order to infer planet masses from dust gaps directly, one typically needs to match observations with customized dusty disk-planet simulations  \citep[e.g.,][]{Dipierro2015OnTau, Clarke2018High-resolutionAu}, which introduces at least two more parameters: the strength of dust-gas coupling (or particle size) and dust abundance (or metallicity). Furthermore, as disk models improve, e.g. with realistic thermodynamics \citep{miranda19,miranda20}, more parameters become necessary to describe planet gaps precisely and place tighter constraints on planet masses. Thus, it will become impractical to perform complex disk-planet simulations to match each observed target in future large surveys of PPDs.

We are therefore motivated to seek a generic,  future-proof procedure to model planet gaps. This calls for a method that can encapsulate the  multi-dimensional parameter space associated with disk-planet interaction. 
In the era of big data, we propose machine learning methods as a solution.

\subsection{A Machine Learning approach}
In this article, we implement an efficient and quick-paced method based on state-of-the-art machine learning (ML) techniques. We build a ML model trained with explicit planet-disk simulations, which only need to be run once and for all. Our ML model can then estimate planet masses given parameters such as the dust gap width, the disk aspect-ratio, disk viscosity, disk surface density profile, metallicity, and particle size.  

The advantage of this ML model is threefold. First, our model can be used by observers as a tool to predict planet mass from observed PPDs \emph{without running any simulations}. 
Second, the model is trained using synthetic data from numerical simulations, hence predicted mass is more accurate (close to the real one) than using  simple empirical relations. {Third, our model can easily be extended to accommodate improved and new physics. This simply amounts to generating new data using advanced simulations and adding relevant variables to the ML algorithm.}

\subsection{Paper plan}

The paper is organized as follows. We first describe the disk-planet systems under consideration in Section \ref{sec:setup}. We then explain, in some detail, the architecture of our artificial neural network in Section \ref{DPNNet_explain} and how the {data is pre-processed and the network is trained to predict planet masses from gap profiles}. We present the predictions from our ML model in Section \ref{sec:floats}, where we also compare its performance to previous empirical approaches. In Section \ref{HL-Tau} we apply it to observed PPDs around HL Tau and AS 209 to infer planet masses. We discuss limitations and future prospects of our basic ML framework in Section \ref{discussion} and conclude in Section \ref{summary}. { Some technical details for the training step are given in the Appendix.}

\section{Disk-planet interaction} \label{sec:setup}

The physical system of interest is a dusty protoplanetary disk with an embedded, gap-opening planet. To model disk-planet interaction using ML, we train the  algorithm with a sample of hydrodynamic simulations. Details of the training step is presented in the next section. Here, we describe the physical disk models, simulations, and notations for later reference.

\subsection{Disk model}
We consider 2D, razor-thin disk models for simplicity. 
Cylindrical co-ordinates $(R,\phi)$ are centered on the central star of mass $M_*$. We place a planet of mass $M_P$ on a Keplerian circular orbit at $R=R_0$ and do not consider planet migration. We adopt a rotating frame with the planet and include the planet's indirect potential. We use dimensionless units such that $G=R_0=M_*=1$, where $G$ is the gravitational constant. We quote time in units of the planet's orbital period $P_0 = 2\pi/\Omega_\mathrm{K0}$, where $\Omega_\mathrm{K} = \sqrt{GM_*/R^3}$ is the Keplerian frequency and sub-script $0$ denotes evaluation at $R=R_0$.  

The disk has an initial gas surface density profile given by: 
\begin{equation}
    \sigg(R) = \Sigma_\mathrm{g0}\left(\frac{R}{R_{0}}\right)^{-\sigma},
\end{equation}
where $\Sigma_\mathrm{g0}=10^{-4}$ (in code units) and $\sigma$ is the exponent of the density  profile. We neglect the disk's gravitational potential as our models have Toomre parameters $Q_0\gtrsim 80$ (see \S\ref{get_data}), which is well-above that required for gravitational stability \citep{toomre64}.

We consider locally isothermal disks with a fixed sound-speed profile $c_s(R) = h_0R\OmK$, where the disk's local aspect-ratio $h_0$ is taken to be a constant, which corresponds to a non-flared disk. The (vertically integrated) gas pressure is then given by $P= c_s^2\sigg$. We include viscous forces to mimic possible turbulence in the gas disk using the standard $\alpha$-prescription \citep{shakura73}, such that the kinematic viscosity $\nu = \alpha c_s^2/\OmK$. We assume constant $\alpha$. 

We also consider a single population of dust particles modeled as a pressureless fluid \citep{jacquet11}, which is valid for Stokes numbers $\st \equiv t_s \OmK\ll 1$, where $t_s$ is the stopping time characterizing the strength of dust-gas coupling \citep{weiden77}. For simplicity, we assume a constant Stokes number, which was found to maintain better numerical equilibrium and behavior near the disk boundaries than a fixed particle-size approach. The dust fluid is initialized with a surface density $\Sigma_\mathrm{d} = \epsilon \sigg$, where $\epsilon$ is the constant (global) dust-to-gas ratio. The reference dust surface density is $\Sigma_\mathrm{d0} = \epsilon \Sigma_\mathrm{g0}$. The backreaction from the dust drag onto the gas is included.

The dusty disk is initialized with steady-state drift solutions given by Equations (70)-(73) of \cite{Benitez-Llambay2019AsymptoticallyFARGO3D}, which assumes a constant Stokes number.

\subsection{Hydrodynamic simulations}
 
We simulate the above dusty disk-planet systems with the \textsc{fargo3d} hydrodynamics code \citep{Benitez-Llambay2016FARGO3D:Code}. \textsc{fargo3d} can be run on Graphics Processing Units (GPUs), achieving large speed-ups compared to Central Processing Units (CPUs). For instance, each of our simulation takes approximately $\sim 12$ GPU hours compared to $\sim 72$ hours on a single CPU core. This advantage, together with the GPU clusters at ASIAA and the National Center for High-Performance Computing in Taiwan, enabled us to run a large number of simulations ($\sim 10^3$) necessary to generate the required training data within a reasonable time.

Our computational domain is $R \in [0.4, 2.5]R_0$ and $\phi\in[0, 2\pi]$, and are discretized with 512 $\times$ 512 uniformly-spaced cells. Periodic boundaries are applied in $\phi$ and the radial boundaries are set to their initial (equilibrium) solutions. We also apply wave-killing zones near the disk edges \citep{valborro06} where the disk is relaxed towards its initial state. The planet is introduced from the beginning of the simulation. We use a constant softening length of $r_s = 0.6h_0R_0$ for the planet's potential.

\begin{figure*}[ht]
\centering

\includegraphics[height=9cm,width=5.0in,trim=55mm 55mm 50mm 30mm, clip=False]{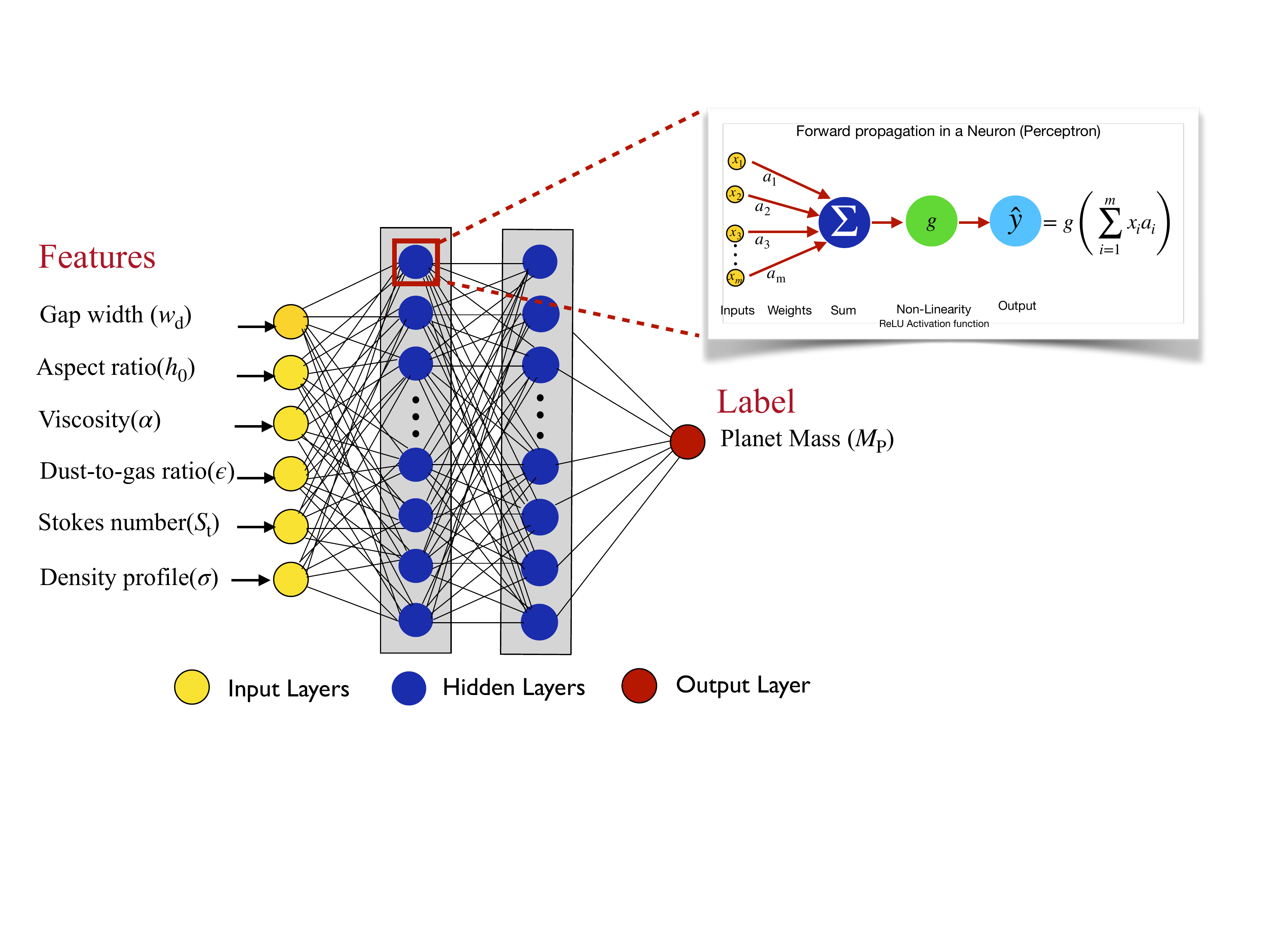}
\caption{The schematic diagram of the {DPNNet using a fully connected multi-layer perceptron}. The first layer in yellow takes as input six feature variables. The output layer in red gives the planet mass as a target variable. The two hidden layers in the middle (in blue) has 128 units of neurons each. Each neuron performs a simple task via forward propagation illustrated in the zoomed panel. It finds the linear combination of inputs (weighted sum) and then passes it through a ReLU activation function to generate an output. The network is trained using RMSprop optimizer. More details in \S\ref{DPNNet_explain}.}
\label{fig:TF_cartoon}
\end{figure*}

\subsection{Parameter space}\label{param_space}

We mainly focus on low mass planets because they are more difficult to observe directly, which makes our ML method more useful \citep{Lodato2019TheDiscs}. 
To this end, we consider planetary masses ranging from $M_P = 8 \, {M_\Earth}$ to $M_P=120\,M_\Earth$ around a $1M_{\odot}$ central star. 
We consider a range of values for disk aspect ratios $h_{\rm 0}\in[0.025,0.1]$, viscosity parameters  $\alpha\in \left[10^{-4}, 10^{-2}\right]$, and power-law slope of the surface density profile $\sigma\in [0.05, 1.2]$. The value of $\sigma$ affects the dimensionless pressure gradient $\eta \equiv \left(h_0^2/2\right)d\ln{P}/d\ln{R} \propto -h_0^2\left(1 + \sigma \right) $, which controls the radial drift of dust \citep{Benitez-Llambay2019AsymptoticallyFARGO3D}.  
For each disk, we consider one dust species characterized by (fixed) Stokes numbers $\st\in[10^{-3}, 10^{-1}]$ and abundance $\epsilon\in[0.01, 0.1]$. 
These values of $S_{\rm t}$ roughly translate to a physical radii of dust particles (or pebbles) ranging from $\sim 1 \rm mm $ to $ \sim 10 \, \rm cm $ assuming an internal grain density of $1 \,\rm g \, cm^{-3} $ and gas surface density $100 \, \rm g \, cm^{-2}$. The parameter space is summarize in Table \ref{tab:parameter space}. 

Each simulation is allowed to evolve for 3000 orbits or about $\sim 1 \rm {Myr}$ at $45 \, \rm au$ around a $1 M_{\odot}$ star. The time is adequate for most models  to reach a quasi-steady state. For some the gaps have not fully settled. However, they are still relevant as they have evolved for an appreciable fraction of the gas disk lifetime.

\begin{table}
\centering
	\caption{Parameter space for hydrodynamic simulations}
	\label{tab:parameter space}
	\begin{tabular}{llll} 
		\hline
		\hline
		Name & Notation & max.  & min. \\
		\hline
		Planet mass in Earth masses & $M_{\rm P}/ M_\Earth$ &  120  & 8 \\
		Disk aspect-ratio &	$ h_{\rm 0}$ &  0.100  & 0.025     \\
		Disk surface density profile & $ \sigma $ &  1.2  & 0.05 \\
        Disk viscosity parameter & $\alpha$ &  $10^{-2}$  & $10^{-4}$  \\
		Global dust-to-gas ratio &$\epsilon$ &  $10^{-1}$  & $10^{-2}$    \\
		Particle Stokes numbers& $ S_{\rm t}$ & $10^{-1}$  & $10^{-3}$ \\ 
		\hline
	\end{tabular}
\end{table}

\section{Implementation of Deep Neural Network}\label{DPNNet_explain}
The last few years have seen a growth in the use of machine learning, a branch of artificial intelligence, in diverse domains to solve complex real-world problems. In particular Deep Learning \citep{Cun15}, a subfield of ML, has been the main driving force behind breakthroughs in image classification \citep{alex12}, speech recognition \citep{Cho15}, text classification \citep{zic16} and more recently self-driven (autonomous) cars \citep{gri20}. Deep Learning allows machines to identify patterns in data using \textit{neural networks} to perform tasks like detection, classification, regression, segmentation, etc. In this work, we are interested in building a neural network and apply it to a \textit{regression} problem \citep[see][]{Goo16,Ali19} in astronomy.   
The task of our neural network is to learn from input variables (also called \textit{features}) like gap-width, disk aspect-ratio, disk viscosity, disk surface density profile, dust abundance, and particle size observed in PPDs to predict a target variable (called \textit{label}) planet mass. In the next section we discuss the structure and the working principle of the network.

\subsection{Setting up the Network}\label{network}
A neural network \citep[see][for a review]{Sch15} consists of multiple layers of neurons or perceptrons. Each neuron performs a simple task via forward propagation, meaning the process of generating output flows in one direction from the input layer to the next layer. A neuron accepts a set of inputs and the corresponding weights. It then passes the linear combination of the inputs through a non-linear activation function to generate an output. The first layer of the network takes as input the features and the last layer gives (predicts) the target variable that is, ideally, close to the true value.  The middle layers are also called the hidden layers, as the states in them are not directly enforced unlike the input and the output layer. The hidden layers accept the outputs from the previous layer and feed their results into the next layer. A network is said to be \textit{deep} when it has more than one hidden layer.
We design a deep neural network, DPNNet (Disk Planet Neural Network) to model planetary gaps using a multi-layer perceptron.

Figure (\ref{fig:TF_cartoon}) gives a schematic representation of our {DPNNet}, with two hidden layers, an input layer (yellow) with six feature variables and an output layer with one target variable. Our network is trained using simulation data to predict the target variable, the planet mass. The training process works iteratively, where the weights of all linear combinations of all neurons are adjusted at each step to get the best agreement between the predicted and the actual value. The matching between the actual and the predicted values is quantified by the cost function, the mean square error (MSE). It is defined as 
\begin{equation}
    C(a) = \frac{1}{N} \sum_{n=1}^{N} (M_{\rm P, predicted,n}-M_{\rm P, simulation,n})^2 ,
\end{equation}
where $N$ is the sample size, $M_{\rm P, simulation,n}$ is the true planet mass used in the simulation, $M_{\rm P, predicted,n}$ is the value predicted by the neural network, and $a$ are the weights. The gradient of the cost function is computed using the chain rule for derivatives, known as back-propagation \citep{Goo16}. The cost is then minimized iteratively using gradient descent \citep{Rud16}. After each iteration, the weights are updated until it reaches an optimal value.

In our DPNNet each of the hidden layers have 128 units of neurons. They are fully connected as each unit of a layer is linked to each unit of the previous and the next layer. We use the ReLU activation function \citep{Nwa18} to introduce non-linearity for each unit. The network is trained using RMSprop optimizer \citep{Rud16} with its parameters set to default values. We select a learning rate of 0.0001. 
The network dimensions are arrived through a series of trial and error tests by varying the number of layers and units per layer. A lower validation loss indicates a better model. Thus we begin by increasing the size of the layers or adding new layers until we get diminishing returns on the validation loss. We further implement L2 regularization \citep{Ng04}, with a coefficient of $0.0001$, and  early stopping to mitigate overfitting.

\subsection{Data acquisition}\label{get_data}
\begin{figure*}[ht]
\centering
\includegraphics[height=5in,width=7.5in,trim=10mm 0mm 0mm 0mm, clip=False]{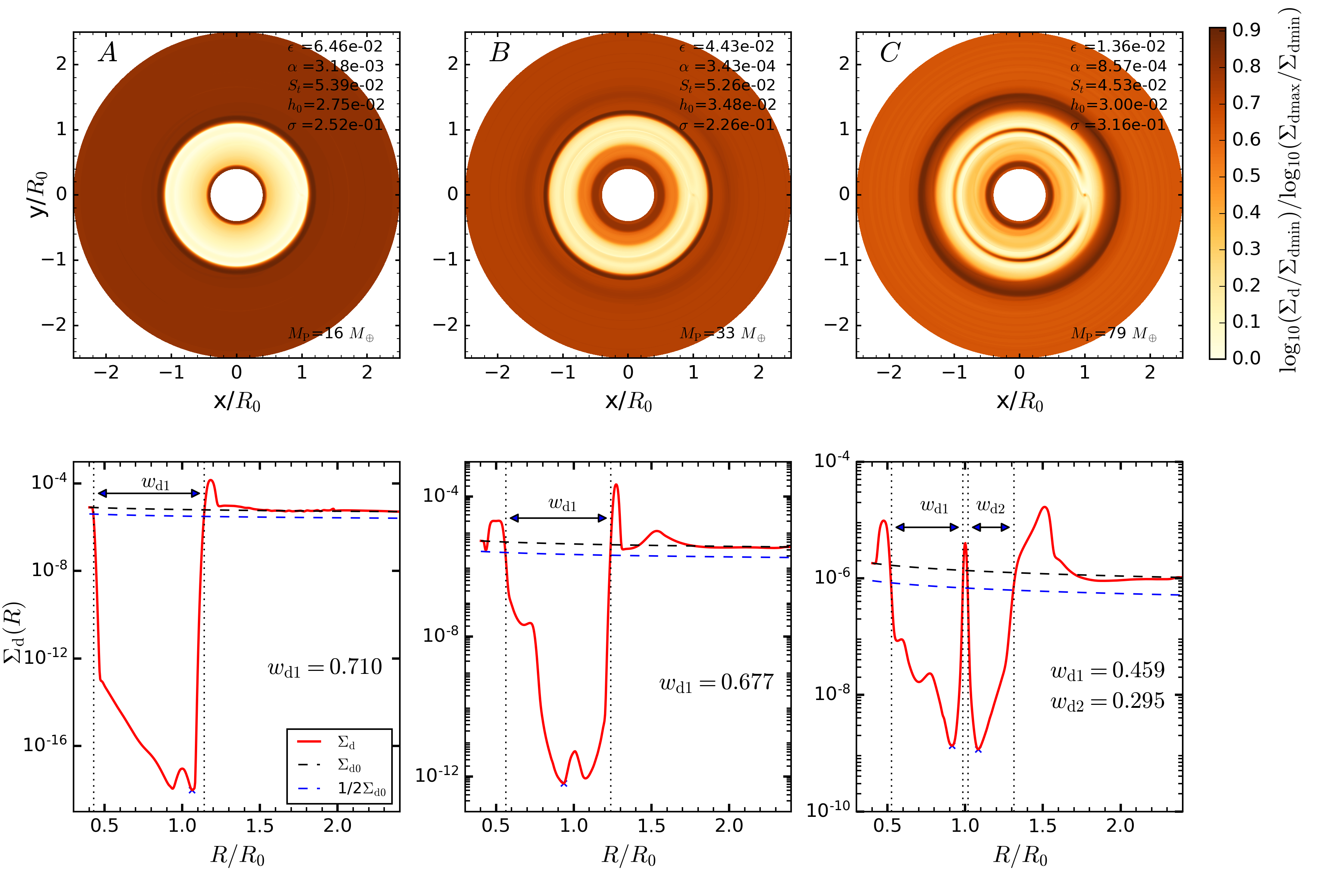}
\caption{The top panel is the normalized dust surface density distribution after $3 \times 10^3$ orbits for different planet masses and different disk initial conditions. The disk parameters for each model are indicated on the top right of each plot. The bottom panel is the radial profile (red line) of the azimuthally averaged surface density for each simulation. The horizontal arrow represents the widths $\wdustone, \wdusttwo$ of the gap(s). The gap width is the distance between the inner and the outer edge of the gap (vertical dotted lines) where $\Sigma_d(R)$ reaches $50 \%$  of the initial surface density $\Sigma_\mathrm{d0}$ shown in dotted blue line. The cross indicates the minimum surface density $\Sigma_{\rm dmin}$.  }
\label{fig:traning_examples_dust}
\end{figure*}

\begin{figure*}
\centering
\includegraphics[height=5in,width=7.5in,trim=10mm 0mm 0mm 0mm, clip=False]{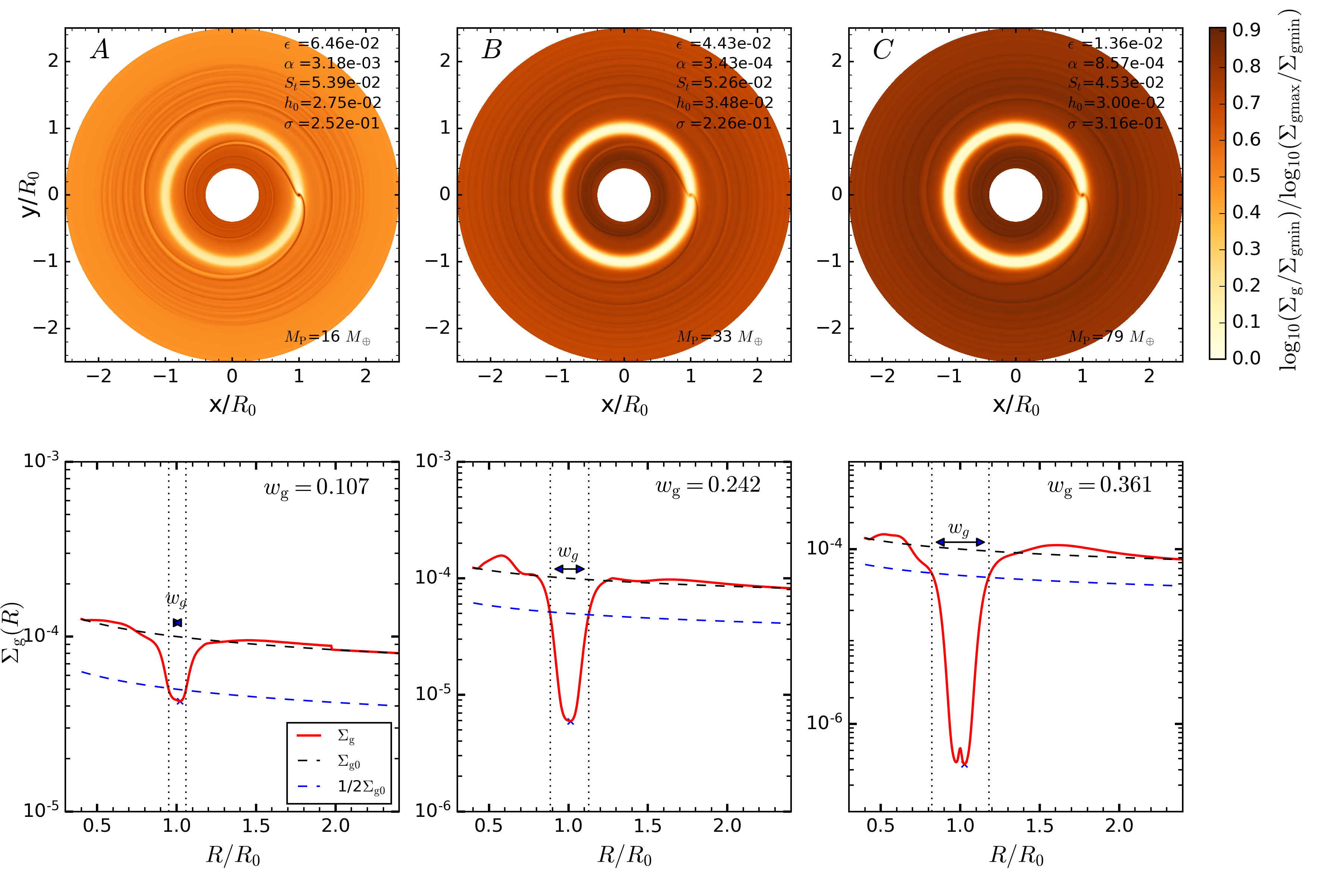}
\caption{Same at Figure \ref{fig:traning_examples_dust} but for gas surface density $\Sigma_{\rm g}$. }
\label{fig:traning_examples_gas}
\end{figure*}

The parameter space for the disk-planet simulations are sampled using Latin hypercube sampling \citep{mck79,ima81} method. We generate a {uniform random distribution of the parameters, with each value centered within the sampling intervals.}  We ran a total of $1100$ simulations across the range of input parameters described in \S\ref{param_space}. Examples simulations are shown in Fig.  \ref{fig:traning_examples_dust} -- \ref{fig:traning_examples_gas} and discussed in more detail below. We analyze the data from each simulation and measure the width of the gap/gaps opened by the embedded planet in both the dust and gas profiles.

Our simulations show that for various parameter combinations, a planet can induce multiple, but adjacent gaps/rings in the dust \citep{dong17}. We thus broadly  classify our simulations into two categories: (a) one deep broad gap as seen in simulation A and B in Figure \ref{fig:traning_examples_dust} and, (b) two gaps adjacent to a planet (see panel C in Figure \ref{fig:traning_examples_dust}). To account for case (b), we simply define two dust gap widths for each simulation, $\wdustone$ and $\wdusttwo$, but set $\wdusttwo = 0$ if only a single dust gap is formed. There are are some simulations that open secondary gaps away from the planet  \citep{bae17,miranda19}. However, we exclude such simulations from our training set as we do not have enough samples to effectively train our models to consider secondary gaps.

Following \cite{Kanagawa2016MassWidth}, we measure the dust gap width as the radial distance between the inner $R_{\rm d, in}$ and outer $R_{\rm d, out}$ edge of the dust gap where the dust surface density $\Sigma_\mathrm{d}$ reaches a predefined threshold fraction of $\Sigma_\mathrm{d0}$, respectively. Thus, the first dust gap width is given by 
\begin{equation}\label{gapwidth}
    w_\mathrm{d1} = \frac{R_{\rm d1,out} - R_{\rm d1, in}}{R_0},
\end{equation}
{where $R_{\rm 0}$ is the planet's orbital radius}, as illustrated in Fig. \ref{fig:traning_examples_dust}. We use similar definitions for the second dust gap width $\wdusttwo$ (if it exists) and the gas gap width $w_\mathrm{g}$ (see Fig. \ref{fig:traning_examples_gas}). For measuring the dust and gas gap widths we adopt a threshold fraction of $1/2 $ of the initial surface density for easy comparison with the Kanagawa model \citep{Kanagawa2016MassWidth}.
One limitation with this definition of $\wdust$ is that it fails for shallow gaps (associated with low mass planets) when the depletion is less than the threshold fraction. However, one can choose any threshold fraction or implement other definition of the gap width. In that case the DPNNet will be trained using those numbers.

Figure \ref{fig:traning_examples_dust} illustrates a subset of simulation outputs from the training set. The top panel shows the dust surface density at the end of $3000$ orbits for three independent runs with distinct planet masses and different disk initial conditions (shown on the upper right corner). Simulation A, B, and C have an embedded planet of mass $M_{\rm P} = 16\, M_\Earth$, $M_{\rm P} = 33\, M_\Earth$ and $M_{\rm P} = 79\, M_\Earth$ respectively. The bottom panel is the corresponding radial profile of the azimuthally averaged surface density. The extent of the dust gap widths $\wdust$ are shown by the horizontal arrows. 

Figure \ref{fig:traning_examples_gas} shows the gas gap surface density and one-dimensional radial profile for the same models as in Figure \ref{fig:traning_examples_dust}. Gas gaps are much smoother and less deep compared to their dust counterpart. The gas gaps mainly fall into category (a), i.e., one gap around the planet, according to the above gap classification scheme. As evident gas gap profiles and features are vastly dissimilar compared to dust gaps.

Figure \ref{fig:traning_examples_dust} highlights the fact that the variation in the gap profiles due to planet-disk interaction is complex and non-linear, which is present even in the simple empirical relations of Eq. \ref{lodato_model}--\ref{eq:kanagawa}. For instance, even though the embedded planet in simulation B is heavier ($\times \, 2.1$) compared to simulation A, they both have comparable dust gap width. This is likely due to the differences in the ambient disk initial conditions, for example simulation B has higher ($\times 1.2$) aspect ratio and lower viscosity ($\times 0.1 $) compared to simulation A.
Furthermore, in simulation C a planet of mass $M_{\rm P} = 79 M_\Earth$ carves a double dust gap of category (b) possibly due to even lower viscosity, $\alpha = 8.57 \times 10^{-4}$, compared to simulation A and B

The initial dataset consisted of $ 1100$ simulations. The data is pre-processed to eliminate runs that do not open up detectable gaps or have secondary gaps. After the initial screening, we have data from $ 976 $ simulations. In addition to the input run parameters ($M_\mathrm{P}, \alpha, h_{\rm 0}, \sigma, S_{\rm t}, \epsilon$), each simulation is also associated with gap width(s) in both dust ($\wdustone$, $\wdusttwo$) and gas ($\wgas$).   
However, the gas gap width will only be used for comparison with other empirical models (\S\ref{modelcomp}), and is \emph{not} part of our ML model.  {Table (\ref{tab:Stat_table}) gives the distribution of the parameters of the raw dataset and the processed dataset after the filtering.}

For building our ML model, we assign each simulation a set of feature variables ($\wdustone, \wdusttwo, \epsilon, \alpha, S_{\rm t},h_{\rm 0},\sigma $) and a target variable (or label), here the planet mass $M_{\rm P}$. {For each feature variable, the data is normalized using the standard scaling (z-score) by removing the mean and then scaling it by the standard deviation.} Note that feature and target variables need not correspond to simulation input parameters and output measurements, respectively. We are free to assign feature and target variables to fit the application, which in our case is to predict planet masses.  It would be straight forward to add, remove, or swap  variables to predict other quantities.

\subsection{DPNNet Training}
The DPNNet is implemented using the Google TensorFlow \citep{tensorflow2015-whitepaper}, which is an open-source platform for machine learning. The data is randomly split into two blocks where $80 \% $ is used for training and validation and the remaining $20 \% $ for testing. This is crucial as we want to test our model on data that it has not previously seen or been trained with. 
The training set is fed into our DPNNet and the weights are adjusted iteratively until they are optimized (see \S\ref{network} for details).{ The validation set is used to check the general performance of DPNNet like avoiding over-fitting. In Figure (\ref{fig:validation_loss}), we plot the training and the validation loss, the MSE, as a function of training epoch for the DPNNet. As evident the validation loss gradually decreases and starts to flatten after 600-700 epoch, whereas the training loss is still decreasing indicating over fitting. Thus we implement early stopping and stop the iterations around 800 epochs. In the next section, we apply our trained network to predict planet mass.}

\begin{figure}
\centering
\includegraphics[width=3.2 in,trim=10mm 0mm 0mm 0mm, clip=False]{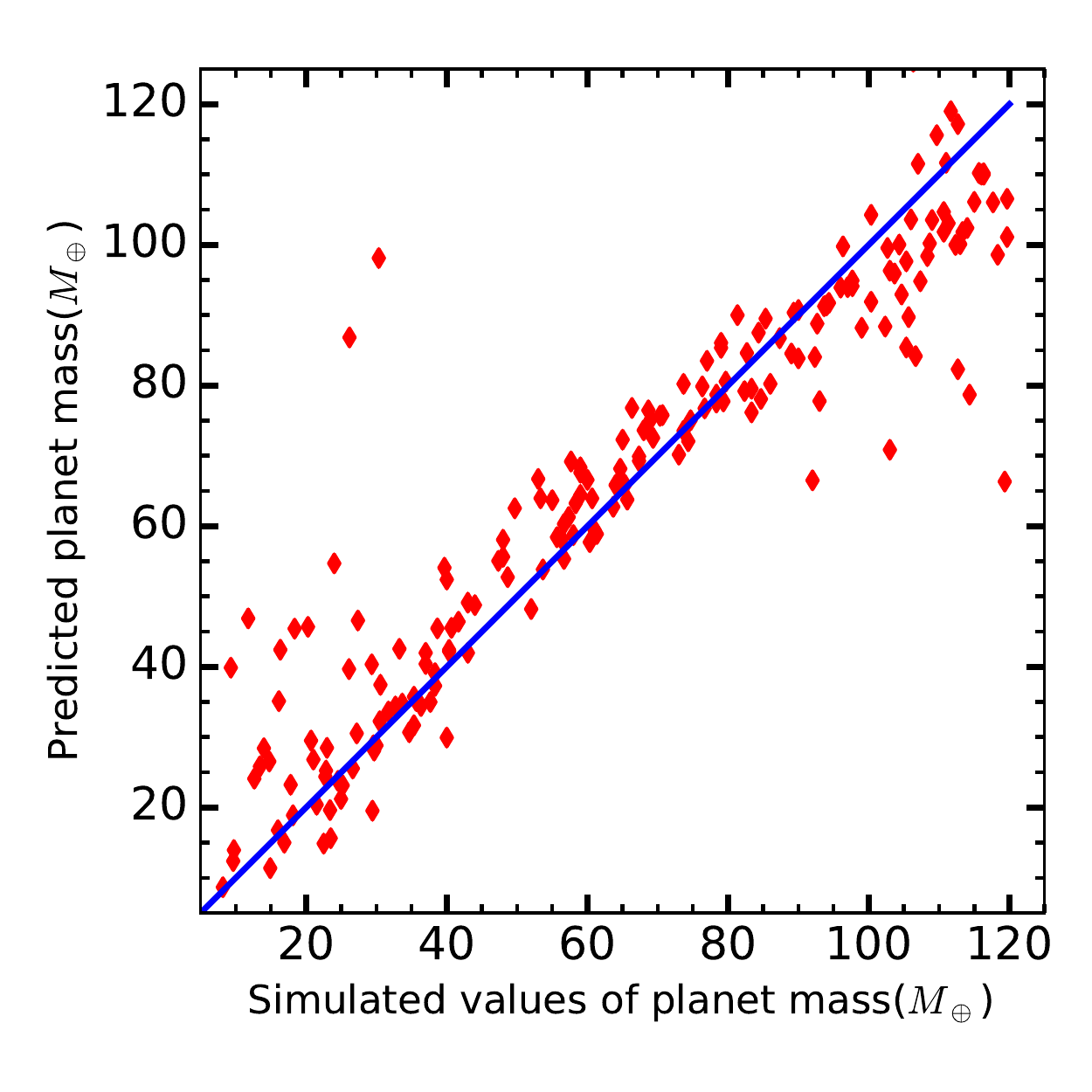}
\caption{Correlation between the predicted planet mass (in the units of $M_\Earth$) using DPNNet and the simulated planet mass used in the numerical simulations.}
\label{fig:prediction}
\end{figure}

\begin{figure}
\centering
\includegraphics[width=3.0 in,trim=10mm 0mm 5mm 0mm, clip=False]{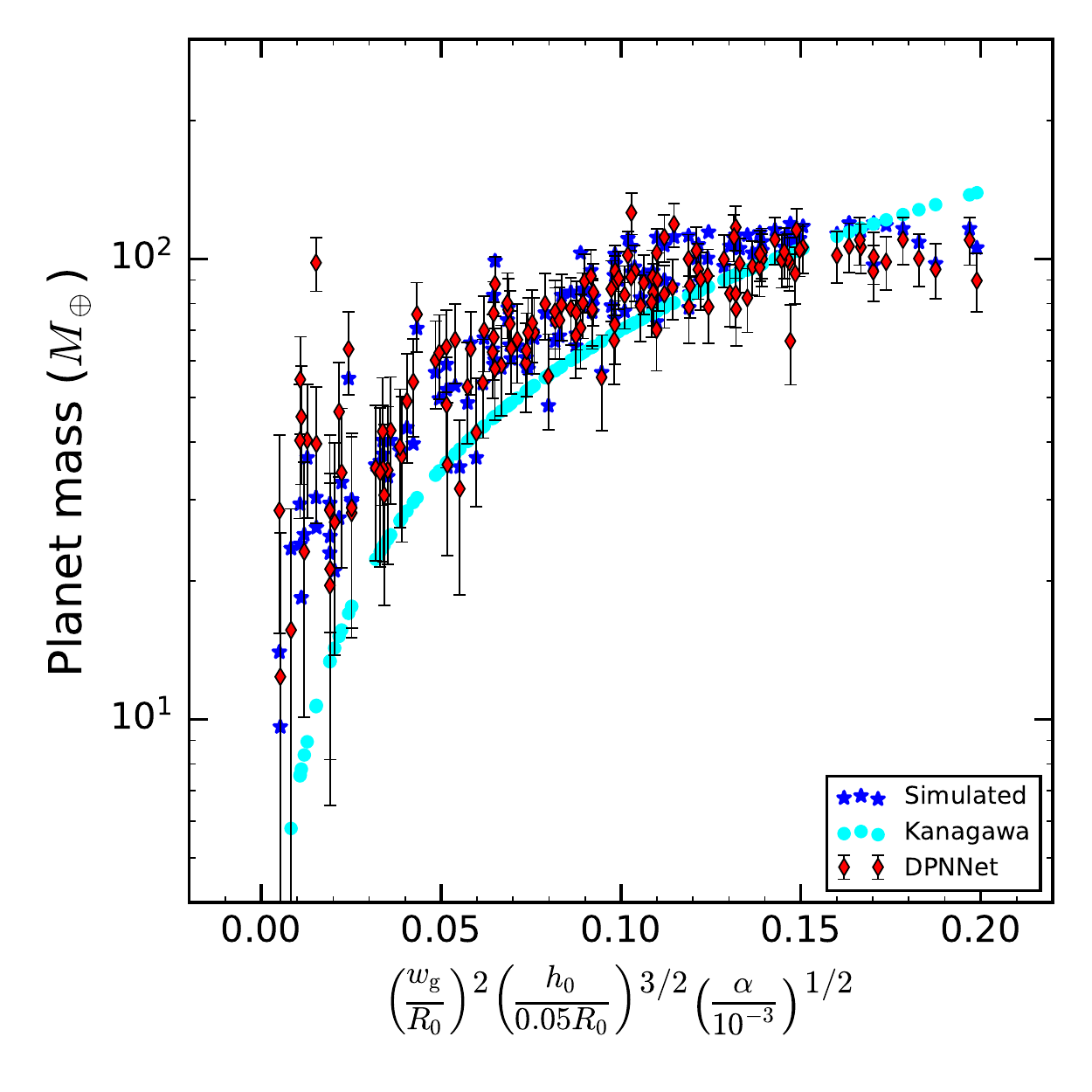}

\caption{Left: Predicted planet mass from DPNNet using $(\alpha, h_{\rm 0}, \sigma, S_{\rm t}, \epsilon, \wdustone, \wdusttwo)$ as input and Kanagawa model using $(\wgas, h_\mathrm{0}, \alpha)$ for input are plotted as a function of gas gap width $w_{\Sigma g}$, aspect-ratio $h_{0}$ and viscosity $\alpha$ as given in Equation (\ref{eq:kanagawa}). The vertical error bars are prediction uncertainty associated with the network (see \S\ref{Model Predcition}). The blue markers are the simulated mass ($M_{\rm P,simulation}$) of the planets. 
}
\label{fig:modelcom1}
\end{figure}

\begin{figure*}
\centering

\includegraphics[height=3.5in,width=7 in,trim=12mm 0mm 6mm 0mm, clip=False]{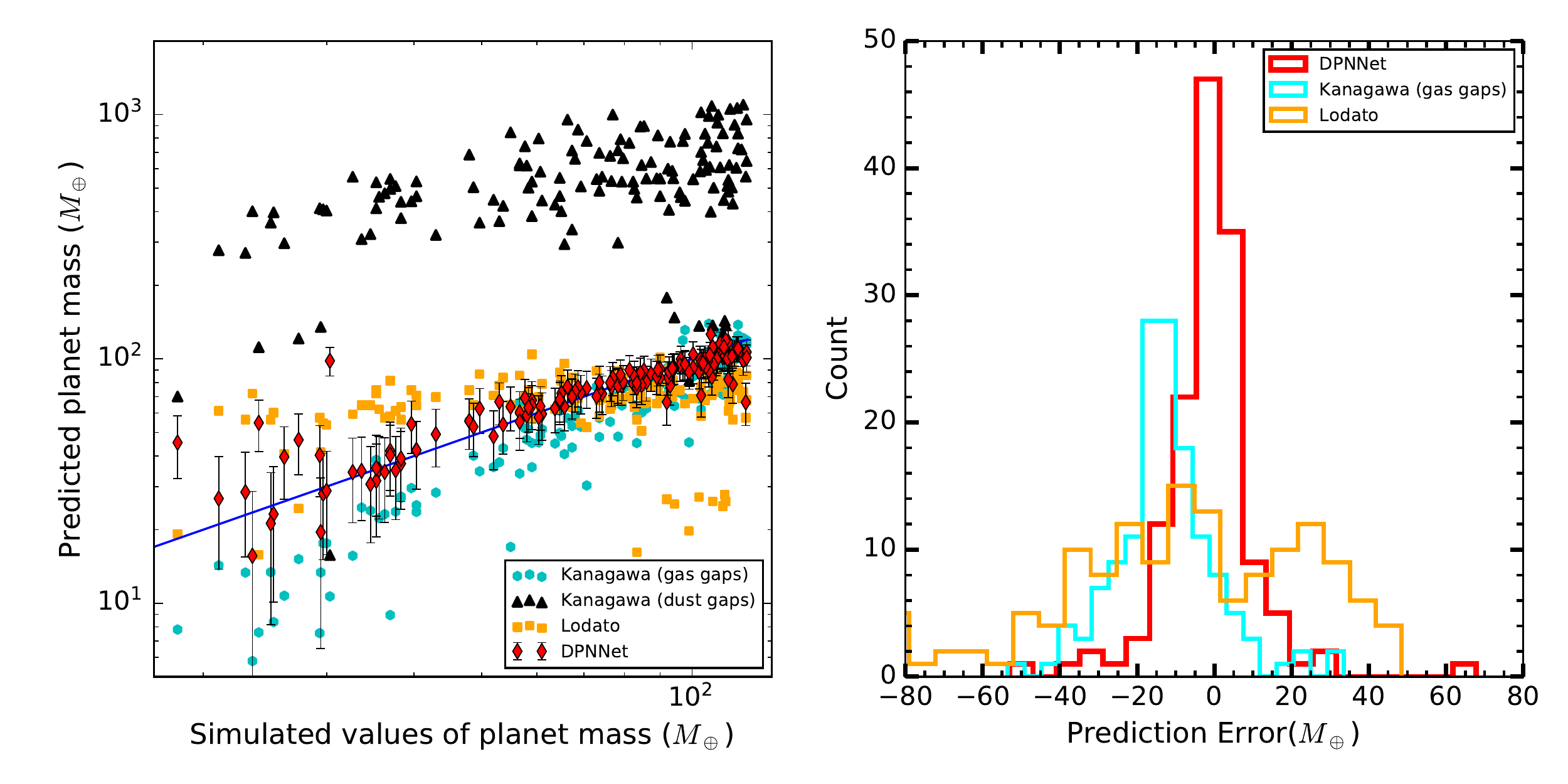}
\caption{Comparison between the planet mass predicted using different models and the simulated planet mass. The DPNNet-predicted mass lies along the blue line indicating close correlation with the true mass. The black triangles are the predicted mass using Kanagawa model from dust gap width, i.e., setting $\wgas\to\wdustone$ in Eq. \ref{eq:kanagawa}.  The Kanagawa models overestimate the planet mass. {The orange squares are the inferred mass from Lodato model (Eq. \ref{lodato_model}). They deviate considerably from the simulated values both in the low and high mass end.}
The masses inferred (cyan hexagon) from kanagawa model using gas gap widths (Eq. \ref{eq:kanagawa} without modification) are marginally below the blue line near the low mass end, indicating under prediction. Right: Distribution of the prediction error ($M_{\rm P,simulation} - M_{\rm P,predicted}$) for Kanagawa, Lodato and DPNNet with ($\mu = - 12.7 M_\Earth$, $\sigma_{\rm E} = 13.1 M_\Earth$), ($\mu = - 7.4 M_\Earth$, $\sigma_{\rm E} = 31.2 M_\Earth$) and ($\mu = - 1.2 M_\Earth$, $\sigma_{\rm E} = 12.0 M_\Earth$) respectively.}
\label{fig:correlation2}
\end{figure*}

\section{Results} \label{sec:floats} 
Once the training process is complete the DPNNet (with optimized weights) is ready to be deployed to predict planet mass from observed disk. We initially test the network's performance by applying it to the test dataset. This allows us to quantify the network's overall accuracy in predicting planet mass from unseen data. We also compare our networks prediction with other empirical relations and discuss its advantages.

\subsection{DPNNet Prediction}\label{Model Predcition} 
The test dataset consists of feature variables along with the true planet mass. {The feature variables are normalized with same distribution that the network has been trained on.
It is then fed to the trained DPNNet to predict the planet mass.} Figure \ref{fig:prediction} illustrates the correlation between the predicted and the simulated planet mass in Earth-mass ($M_\Earth$) units for all the test samples. The predicted mass (red diamonds) lies along the blue line indicating a close correlation. We estimate the error ($M_{\rm P,simulation} - M_{\rm P,predicted}$) between the predicted and the simulated mass. It is normally distributed with a mean ($\mu$) and a standard deviation ($\sigma_{\rm E}$) of $0.8 \, M_\Earth$ and  $12.5\, M_\Earth$ respectively. The model performance is evaluated using standard metrics such as the root mean square error (RMSE) and the mean absolute error (MAE). The DPNNet has a RMSE of $12.5 \, M_\Earth$ and a MAE of $ 7.9 \, M_\Earth$ when applied to the test dataset. 

Furthermore, to lower the bias associated with data sampling we apply k-Fold cross validation resampling process using scikit-learn \citep{scikit-learn}. Instead of using a single RMSE we take the average of the RMSE obtained from each of the 5-fold resampling process. We get a mean RMSE of $ 13.0 M_\Earth$. This can be considered as the prediction uncertainty of our network. Given our limited sample size and wide parameter space the model error is moderate, particularly towards the high mass end of our parameter space. The network performance can be further improved by using more comprehensive data. However, this is beyond the scope of this work.  

In the next section, we compare DPNNet with both the Kanagawa \citep{Kanagawa2016MassWidth} and the Lodato model \citep{Lodato2019TheDiscs}.

\subsection{Model Comparison} \label{modelcomp}
We apply the Kanagawa and Lodato models along with DPNNet to the test dataset to generate planet masses and compare it with simulated values. This is done as follows. 
Each run in the test dataset is associated with some combination of 
$(M_\mathrm{P}, \alpha, h_{\rm 0}, \sigma, S_{\rm t}, \epsilon, \wdustone, \wdusttwo, \wgas)$. This $M_\mathrm{P}$ is the simulated planet mass. The Lodato model (Eq. \ref{lodato_model}) takes a dust gap width as the sole input to predict a planet mass, in which case we use $\wdustone$. Similarly, we only use
$(\wgas, h_\mathrm{0}, \alpha)$ for input in the Kanagawa prediction (Eq. \ref{eq:kanagawa}). Finally, our DPNNet takes $(\alpha, h_{\rm 0}, \sigma, S_{\rm t}, \epsilon, \wdustone, \wdusttwo)$ as input to predict a planet mass.

{Figure \ref{fig:modelcom1} shows the simulated planet mass (blue star) and predicted planet mass using the Kanagawa model\footnote{For applying the Kanagawa model we select only those simulations that have a measurable gas gap.} (cyan hexagons). 
The Kanagawa planet masses are comparable to the simulated values with a RMSE of $18.2 M_\Earth$. While the Kanagawa model perform reasonably well towards the higher mass end, one caveat is that it requires gap widths measured in gas emission. This may limit its application, as observed gap profiles are mostly from dust emission. One would then first need to estimate gas gap widths, which can introduce large uncertainties \citep{Zhang2018TheInterpretation}. }

On the other hand, our DPNNet takes as input dust gap widths (a more direct observable) along with other disk features. The DPNNet-predicted planet masses are indicated in red diamonds in Figure \ref{fig:modelcom1}. The vertical error bars represent the uncertainty ($\pm 13.0 M_\Earth$) in the predicted mass. The RMSE for the DPNNet is $12.1 M_\Earth$. Our DPNNet predictions have much reduced bias and produce better fits compared to the Kanagawa model.

Figure \ref{fig:correlation2} makes a more direct comparison between models. It shows the predicted $M_{\rm P}$ from dust gap widths and disk parameters as a function of actual planet mass. An exact prediction would lie on the blue line. 
The black triangles are the predicted mass from the Kanagawa model, but using the dust gap width instead of the gas gap width (setting $\wgas\to\wdustone$ in Eq. \ref{eq:kanagawa}). These are an order of magnitude above the blue line, which shows that using dust gap widths in the Kanagawa models lead to significant overestimates in planet mass. The orange squares are the inferred planet mass using the Lodato model . {We re-scale the proportionality constant $k$ in Eq. (\ref{lodato_model}) to adjust for the different way we measure the gap width compared to \cite{Lodato2019TheDiscs}. We fit the measured dust gap width $\wdustone$ to $k \times (M_{\rm  P}/3M_{*})^{1/3}$, where $M_{\rm P,*}$ are the simulated planet and stellar mass, using the training dataset and obtain $k=18.4$. The estimated masses deviate considerably from actual simulated values both in the low and high mass end. The orange histogram on the right denotes the distribution of the error in the Lodato model. It follows a normal distribution with $\mu = -7.4 M_\Earth$ and $\sigma_{\rm E} = 31.2 M_\Earth$.} 

{On the other hand, the Kanagawa model using gas gap widths (cyan hexagons; Eq. \ref{eq:kanagawa} without modification) typically underestimates planet masses, although it performs better in the high mass end. The cyan histogram on the right follows a normal distribution with $\mu = -12.7 M_\Earth$ and $\sigma_{\rm E} = 13.1 M_\Earth$. The peak of the histogram is considerably shifted from the center indicating a negative bias in its estimate.
In contrast, the mass predicted using the DPNNet lies along the blue line indicating a much closer correlation than other models. Thus, our DPNNet outperforms both empirical relations when applied to the test dataset. The error follows a normal distribution (red histogram) with $\mu = -1.2 M_\Earth$ and $\sigma_{\rm E} = 12.0 M_\Earth$.}

\section{Application to observations}\label{HL-Tau}

As a first application of our newly developed DPNNet, for planet gaps, we 
estimate planet masses responsible for opening the dust gaps observed in the  protoplanetary disks around HL Tau and AS 209. Our findings are summarized in Table \ref{tab:HL-Tau} and discussed below.

\begin{table*}

\centering
\caption{Inferred planet masses from gaps in HL Tau and AS 209}
\begin{tabular}{|l|l|l|l|l|l|l|l|l|l|l|l|l|l|l|l|}

 \hline
\multicolumn{9}{|c|}{Observed Features} &\multicolumn{1}{|c|}{DPNNet} & \multicolumn{6}{|c|}{Other models}\\
 \hline
  Name & $M_{*}$ & $R_{\rm gap}$  & $w_\mathrm{d1}$ & $h_{0}$  & $\alpha$ & $S_{\rm t}$ & $\epsilon$ & $\sigma_{0}$ & $M_{\rm p0}$ & $M_{\rm p1}$ & $M_{\rm p2}$ & $M_{\rm p3}$ & $M_{\rm p4}$ &$M_{\rm p5}$ & $M_{\rm p6}$ \\
  & ($M_{\odot}$) & (au)  &  &  &  & &  &  & ($M_{\rm J}$)  & ($M_{\rm J}$) & ($M_{\rm J}$) & ($M_{\rm J}$) & ($M_{\rm J}$) & ($M_{\rm J}$)& ($M_{\rm J}$) \\
  &  &  &  &  &  & &  &  & $\pm 0.04 M_{\rm J}$  &  & & & && \\
 \hline

		\hline
		               & 1.00                           & 10     &  0.81  & 0.05  & $10^{-3}$ & 0.005  &  0.01 & 1.0 & 0.25 & 1.40 & 0.20 &0.35 & 0.20  & - & - \\
		HL Tau         & 1.00                           & 30     &  0.23  & 0.07  & $10^{-3}$ & 0.005  &  0.01 & 1.0 & 0.20 & 0.20 & 0.20  &0.17 & 0.27 & - & - \\
		               & 1.00                           & 80     &  0.29  & 0.10  & $10^{-3}$ & 0.005  &  0.01 & 1.0 & 0.22 & 0.50 & 0.20 &0.26 & 0.55. & - & -  \\
		                
				\hline
				
		AS 209         & 0.83                           & 9       &  0.42    & 0.04  & $10^{-4}$ & 0.016  &  0.01 & 1.0 & 0.13 & - & - & - & - & 0.37 &  -   \\
		               & 0.83                           & 99      &  0.31    & 0.08  & $10^{-4}$ & 0.016  &  0.02 & 1.0 & 0.14 & - & - & - & - & 0.18 & 0.21  \\
		\hline
	\end{tabular}
	\tablecomments{$M_{\rm P0}$ is the inferred mass using DPNNet from input features $(\wdustone, h_{\rm 0},\alpha,S_{\rm t}, \epsilon, \sigma )$ given in column 4 - 9  and setting $\wdusttwo=0$.   $M_{\rm P1}$ is the mass predicted by \cite{Kanagawa2016MassWidth} using the empirical relation. $M_{\rm P2}$ ,$M_{\rm P3}$, and $M_{\rm P4}$ are the masses obtained using customized simulation of HL Tau by \cite{Dong2015OBSERVATIONALDISKS, Jin2016MODELINGINTERACTIONS} and \cite{ Dipierro2015OnTau} respectively. $M_{\rm P5}$ are the inferred planet mass in AS 209 by \cite{Zhang2018TheInterpretation}. $M_{\rm P6}$ is planet mass from \cite{fed18} using specialized simulations of AS 209.}\label{tab:HL-Tau}

\end{table*}

\subsection{HL Tau} 

We apply our DPNNet to the HL Tau disk, which shows clear evidence of axisymmetric gaps in dust thermal emission \citep{ALMA15}. Several studies using hydrodynamic simulations suggest such rings are due to disk-planet interaction \citep{Dong2015OBSERVATIONALDISKS, Dipierro2015OnTau, Jin2016MODELINGINTERACTIONS}. Assuming that each gap is created by a single planet, we use the properties of the observed gap as input features (see Table \ref{tab:HL-Tau}) for our ML model to predict the planet mass responsible for carving out each gap. 
For instance, the disk aspect-ratios and gap widths for three identified gaps at $R= 10 \, \rm au$, 30 au and 80 au in HL Tau are obtained from \cite{Kanagawa2015FormationRotation,Kanagawa2016MassWidth}. Those were estimated for an assumed spectral index $\beta = 1.5$, using optical depth and the gas temperature from the brightness temperatures in ALMA Band 6 and 7. We set the viscosity $\alpha= 10^{-3}$ \citep{Dipierro2015OnTau}, a central star mass $M_{*}= 1 M_{\odot}$, and a canonical value of the dust-to-gas ratio $\epsilon = 0.01$ \citep{ALMA15,Dong2015OBSERVATIONALDISKS}. The Stokes number is of the order of $10^{-2}-10^{-3}$ around the three gaps \citep{Dong2015OBSERVATIONALDISKS}. For simplicity we adopt $S_{t} = 0.005$. Our DPNNet predicts the planet mass in gaps at $10$ au, $30$ au, and $80$ au as $80 \, M_\Earth$,  $63 \, M_\Earth$, and $70 \, M_\Earth$ with an uncertainty of $\pm 13 \, M_\Earth$. These masses correspond to $0.25M_J$, $0.20M_J$, and $0.22M_J$. 

These estimates are in excellent agreement with the inferred planet masses from direct numerical simulations. 

For instance, \cite{Dong2015OBSERVATIONALDISKS} and \cite{Jin2016MODELINGINTERACTIONS} generated synthetic images using three-dimensional Monte Carlo Radiative Transfer and hydrodynamical simulations of disk-planet systems to compare with observations. \cite{Dong2015OBSERVATIONALDISKS} established that three planets each of mass $0.2 M_{\rm J}$ were responsible for the gaps in HL Tau. \cite{Jin2016MODELINGINTERACTIONS} found three embedded planets with masses $0.35 \, M_{\rm J}$, $0.17 \, M_{\rm J}$, and $0.26 \, M_{\rm J}$ in the gaps. Furthermore, three-dimensional dusty smoothed particle hydrodynamics calculations by \cite{Dipierro2015OnTau} yield comparable masses of $0.20 \, M_{\rm J}$, $0.27 \, M_{\rm J}$ in the two inner gaps and a larger $0.55 \, M_{\rm J}$ mass planet in the outer gap.
However, these inferred planet masses (from our DPNNet and specialized simulations) particularly at the inner gap differ considerably with the Kanagawa model \citep{Kanagawa2016MassWidth}. They predicted planet masses of $1.40 \, M_{\rm J}$, $0.20 \, M_{\rm J}$, and $0.5 \, M_{\rm J}$ at $10$ au, $30$ au, and $80$ au respectively using the dust gap widths.
This may be because the Kanagawa model often over-estimates planet mass when applied to dust gap width (see \S\ref{modelcomp}).

\subsection{AS 209}
AS 209 is another interesting system with multiple gaps at 9, 24, 35, 61, 90, 105 and 137 au \citep{Hua18b,guz18} surrounding a central star of mass $M_{*} = 0.83 M_{\odot}$. \cite{Zhang2018TheInterpretation} established using numerical simulation that a single planet of mass $\sim 27 M_\Earth $ $(M_\mathrm{P} = 10^{-4}M_*)$ at $R \sim 100$ au in a $\alpha \simeq 10^{-5}$ disk (with radially varying $\alpha$) can produce all the five gaps at 24, 35, 62, 90, and 105 au. The disk aspect-ratio $h_{0}\sim 0.05-0.06$ was constrained using the distance between two dominant gaps at $R= 61$ and $100$ au. \cite{Zhang2018TheInterpretation} also suggested that the innermost gap at $9$ au hosts a second planet.
Furthermore, \citet{fed18} used customised 3D hydrodynamical simulations of planet-disk interaction to establish that a single planet of $M_{\rm P} \sim 0.7 M_{\rm Saturn}$ ($\sim 67 M_{\Earth} $) at $R \sim 103$ can open up the two gaps detected  at 62 and 103 au.

However, the DPNNet is not explicitly trained for systems where multiple ($>2$) gaps are induced by a single planet. Thus, for simplicity, when applying to AS 209 we only consider the planet-harboring gaps at $R = 9 $ and $99$ au and ignore the others. The potential effect on the predicted mass due to this approximation is likely to be within the model uncertainly. Advance application considering the influence of these subsidiary gaps will be implemented in future improvements.

As input to our network we adopt disk parameters and dust gap widths\footnote{Note that \cite{Zhang2018TheInterpretation} define the gap width differently than us. This is likely to introduce additional uncertainty in the predicted planet mass.} from \cite{Zhang2018TheInterpretation}. While \cite{Zhang2018TheInterpretation} include a range of values for dust abundance, particle size and disk viscosity, we consider those which are within our parameter space (Table \ref{tab:parameter space}). For example, we select $S_{t} = 30 \times 5.23 \times 10^{-4}$ and $\alpha = 10^{-4}$ for both the gaps. 
Next, for the innermost gap at $R = 9 $ au, the other selected parameters are $\wdustone = 0.42, \epsilon = 0.012, \, \rm {and} \, h_{0} = 0.04$. While for the gap at $R = 99$ au, we have $\wdustone = 0.31, \epsilon = 0.017 \, \rm {and} \,  h_{0} = 0.08$.
We set surface density profile $\sigma= 1$ \citep{fed18} for both the gaps. The feature variables are summarized in Table \ref{tab:HL-Tau}. 

The DPNNet predicts a planet mass of $39M_{\Earth}$ and $45 M_{\Earth}$ ($\pm 13 \, M_\Earth$) in 9 au and 99 au gaps respectively. These masses correspond to $0.13M_{\rm J}$ and $0.14M_{\rm J}$. Our estimates, particularly at $R= 9$ au, are lower compared to  \citet{Zhang2018TheInterpretation} values, as they find masses of $0.37 \, M_{\rm J}$ and $0.18 \, M_{\rm J}$ at 9 au and 99 au respectively for the same disk parameters. The discrepancy could be partially because \citet{Zhang2018TheInterpretation} measured gap widths differently than us and for not considering the effects of subsidiary gaps. Also the disk parameters adopted from them are likely to be uncertain.

Given the simplicity of our setup, limited parameter range and sample size used in our training set, our DPNNet performs remarkably well as its predictions are consistent with specialized models and simulations, which were exclusively customized to reproduce selected observations. 

\section{Discussion}\label{discussion}

We introduce an ingenious multi-disciplinary technique that combines computational astrophysics with deep learning (using neural networks) to characterize unseen exoplanets from observed disks. Deep neural network 
learns the underlying relationship in a set of data via training. In our case, it maps the connection between the planetary gaps in dust and the complex disk-planet interaction. This enables our DPNNet to directly estimate planet masses from observed dust gap width and disk features.

Previous analytic scaling relations that can infer planet mass from gas surface density are often limited as PPDs are mostly detected in dust emission. Using these empirical relations requires one to estimate gap profiles in gas from dust continuum flux, which is uncertain and prone to large errors \citep{Zhang2018TheInterpretation}. While customized disk-planet simulations for modeling specific PPDs have been successful, this approach may not be efficient in dealing with multiple targets in a large survey. Thus machine learning, more specifically deep learning, provides an alternative way that is much more accurate and definitive in characterizing unseen exoplanets from observations of PPDs. 

The key to a well-trained network is a comprehensive dataset that reflects the underlying connection between the input features and the output (target) variables.
This necessitates the need to include data that incorporates not only a broad parameter space but also captures the diverse disk morphologies due to disk-planet interaction. However, as the first paper in a series, we have only explore a limited range due to finite resources. This restricts the scope of our network as discussed below.

\subsection{Network Limitations}
The network is trained with limited data generated using  idealized 2D hydrodynamical simulations. Our limited sample size for the training set affects the network performance resulting in higher uncertainty in its prediction. For example, our optimized network has an uncertainty of $\pm 13.0 M_\Earth$, which corresponds to $11 \%$ error even for the most massive planet in our parameter space. The relative uncertainty is much higher at the low mass end. Runs resulting in complex morphologies such as secondary/tertiary gaps have to be excluded as we do not have enough samples to successfully train the network to identify them.

Furthermore, we only consider gaps opened by a single planet in a narrow mass range ($8M_\Earth\le M_{\rm P} \le 120 M_\Earth$). These allow for smoother gap profiles, which are easier to characterize, compared to massive ones as those are likely to induce eccentric gaps \citep{pap02,kle06} and have vortices at gap edges \citep{Kol03,li05,de07} . This further prevents the application of the network to disks harboring massive planets.

We only measure the gap widths ($\wdust$) after 3000 orbits, whence the majority of the simulations have settled into a quasi-steady state. 
Here, an improved approach would be to measure the gap profiles as a function of time as well, such that our network learns to identify the variations due to time evolution in the observed gap. This would then allow one to relax the assumption that the observed system has reached a steady state, and use the system age as another input parameter. 

Our idealized 2D simulations can also be improved. For instance, a fixed particle-size approach in place of a constant Stokes number, adopted here for better numerical behavior, would be a more physical model for dust-gas interaction \citep{weiden77}. 

{ Previous work showed that the vertically-integrated gap structure in gas is similar between 2D and three-dimensional (3D) simulations \citep{fung16}. Furthermore, solids tend to settle into a thin dust layer around the disk midplane. Thus, we do not expect the relation between dust gap profiles and planet masses, as identified by our DPNNet based on 2D simulations, to differ significantly from that trained with 3D simulations. Nevertheless, 3D disk models should be considered in the future as these can also account for disk and planet inclinations, which cannot be captured in 2D. At present the cost of 3D simulations likely prohibits one to obtain a sufficiently large training set. However, if this can be overcome then one only needs to re-train our DPNNet for application to 3D disks.  
}

\subsection{Future prospects}
Our DPNNet demonstrates a simple but powerful application of deep learning in constraining the mass of hidden exoplanets. With more realistic simulations and a larger training dataset, the network can possibly be used to characterize additional properties like gas surface densities and gap depths from observed PPDs. Improved and new physics can be accommodated by adding more variables to the training set as far as the neural network is concerned. In the future, with the emergence of powerful computers, it will be possible to have adequate data to train the network with synthetic images generated from 3D radiative transfer and hydrodynamical simulations, allowing fast and direct comparisons between models and observations. 
This will potentially make the network versatile enough to entirely replace hydrodynamical simulations for observers and modelers.

\section{Conclusions}\label{summary}

In this paper we design deep neural network, DPNNet to estimate the mass of a planet from the dust gap it opens in a protoplanetary disk. The important takeaway are summarized as follows:

\begin{itemize}

\item The DPNNet can directly infer planet mass from gap width observed in dust emission and other disk features. It takes as input dust gap widths, gas disk properties (aspect-ratio, viscosity, surface density profile), and dust properties (abundance, Stokes numbers) from observations and estimates the embedded planet mass. 

\item Compared to previous empirical models \citep[e.g.,][]{Kanagawa2016MassWidth,Dong2015OBSERVATIONALDISKS,Lodato2019TheDiscs}, the DPNNet additionally accounts for changes in the predicted planet mass due to variation in dust abundance, Stokes number, and disk surface density profile.

\item The DPNNet is applied to a test dataset to evaluate its effectiveness. The DPNNet-predicted mass is closely correlated with the true planet mass. Our current model uncertainty is $13 M_\Earth$, but this can be improved in the future with a larger training set. 

\item We deploy our DPNNet to infer planet masses from the dust gaps observed in the protoplanetary disks around HL Tau and AS 209. We find planet masses of $80 \, M_\Earth$, $ 63\, M_\Earth$ and $70 \, M_\Earth$ in gaps at 10au, 30 au, and 80 au, respectively, in the HL Tau disk. Similarly, for the AS 209 disk we infer embedded planet masses of $ 39\, M_\Earth$ and $45 \, M_\Earth$ at the 9 au and 100 au gaps, respectively. These estimates are in close agreement with results from other models based on specialized disk-planet simulations. The result is summarized in Table \ref{tab:HL-Tau}. 
\end{itemize}

\acknowledgments

{We thank the anonymous referee for the constructive comments.} This work is supported by the Ministry of Science and Technology of Taiwan (grant number 107-2112-M-001 -043-MY3). Numerical simulations were performed on the TIARA cluster at ASIAA, as well as the TWCC cluster at the National Center for High-performance Computing (NCHC). We are grateful to the NCHC for computing time,  facilities, and support.

\newpage

\begin{figure}
\centering
\includegraphics[width=3.1 in,trim=10mm 0mm 0mm 0mm, clip=False]{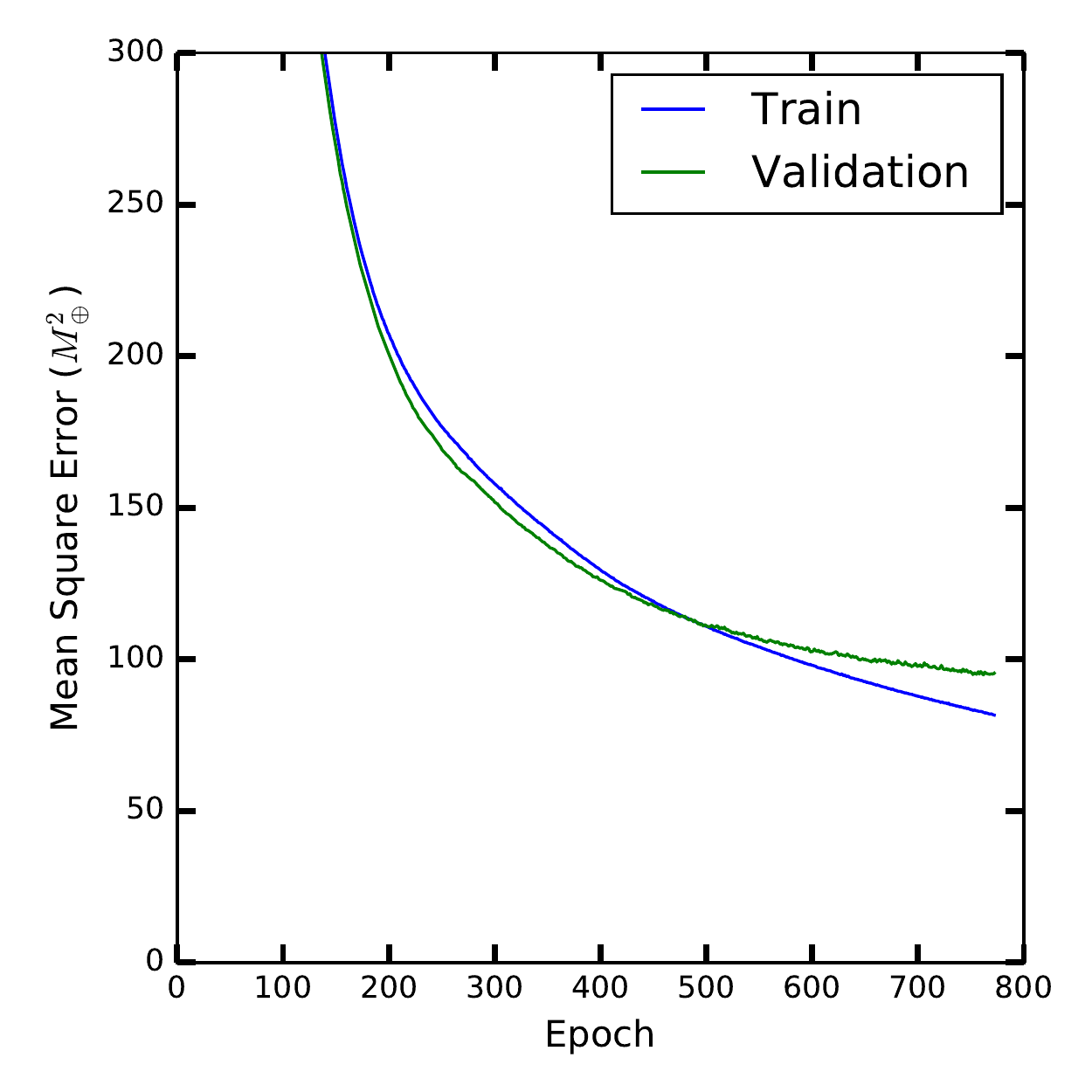}
\caption{Training and validation loss as a function of training epoch for the DPNNet for computation of the planet mass.}
\label{fig:validation_loss}
\end{figure}

{ 
\appendix

Table (\ref{tab:Stat_table}) gives the statistical distribution of the simulation dataset before and after the screening. The raw dataset corresponds to the data from the entire 1100 simulations. The filtered dataset is obtained after the initial cut to remove runs that do not open up detectable gaps or have more that two gaps. While the variables ($M_\mathrm{P}, \alpha, h_{\rm 0}, \sigma, S_{\rm t}, \epsilon$) are the input parameters for the simulations, the dust gaps width are measured from the dust density profile from the simulation outputs.
Figure (\ref{fig:validation_loss}) shows training and the validation loss, the MSE, as a function of training epoch for the DPNNet. As evident the validation loss gradually decreases and starts to flatten after 600-700 epoch, whereas the training loss is still decreasing indicating over fitting. We implement early stopping and stop the iterations around 800 epochs.
}

\begin{table*}

\centering
\caption{Distribution statistics of parameters for the simulation data}
\begin{tabular}{|l|l|l|l|l|l|l|l|l|l|}
 \hline
 \multicolumn{2}{|c|}{}&\multicolumn{4}{|c|}{Raw data} & \multicolumn{4}{|c|}{Filtered data}\\
 \hline
  Name & Notation  & max.  & min.  & mean & std   & max.  & min.  & mean & std\\
  
 \hline

		\hline
		Planet mass in Earth masses & $M_{\rm P}/ M_\Earth$& 120 .0 & 8.1 & 64.2 & 32.4  & 120.0 &  8.2 & 66.8 &31.5 \\
		Disk aspect-ratio &	$ h_{\rm 0}$  &  0.099  & 0.025 & 0.040 & 0.011  &  0.091  & 0.025 & 0.038 & 0.009 \\
		Disk surface density profile & $ \sigma $  &  1.20  & 0.05 & 0.54& 0.29  &  1.20  & 0.05 & 0.53 & 0.29\\
        Disk viscosity parameter & $\alpha$ ($\times 10^{-3}$) &  10 & 0.11  & 5.03 & 2.86  &  10  & 0.13 & 5.13 & 2.77 \\
		Global dust-to-gas ratio &$\epsilon$ & 0.10  & 0.01 & 0.05 & 0.03  & 0.10  & 0.01 & 0.05 & 0.03\\
		Particle Stokes numbers& $ S_{\rm t}$  &  0.100  & 0.001 & 0.051 & 0.029&  0.10  & 0.001 & 0.053 & 0.028\\ 
		Dust Gap one & $ \wdustone $  &  0.94  & 0.00 & 0.67 & 0.22  &  0.94  & 0.10 & 0.74 & 0.11\\
		Dust Gap two& $ \wdusttwo $  &  0.69  & 0.00 & 0.03 & 0.09  &  0.47  & 0.00 & 0.02 & 0.07 \\

		\hline
	\end{tabular}
	\tablecomments{Raw data ($N=1100$) corresponds to the simulation dataset. Filtered data ($N=976$) is obtained after the initial screening (see \S\ref{get_data} for details).  }
\label{tab:Stat_table}

\end{table*}

\bibliography{references_mod}{}
\bibliographystyle{aasjournal}

\end{document}